\documentclass[journal]{IEEEtran}
\usepackage{amsmath,amssymb,amsfonts}
\usepackage[caption=false,font=normalsize,labelfont=sf,textfont=sf]{subfig}
\usepackage{color}
\usepackage{stfloats}
\usepackage{verbatim}
\usepackage{graphicx}
\usepackage{bm}
\usepackage{cite}
\newtheorem{theorem}{\bf Theorem}
\usepackage{booktabs}
\usepackage{enumitem}
\usepackage{cases}
\usepackage[]{hyperref}

\usepackage[linesnumbered, ruled]{algorithm2e}
\SetKwRepeat{Do}{do}{while}%

\begin{document}

\title{Distributionally Robust Game for Proof-of-Work Blockchain Mining Under Resource Uncertainties}
\author{Xunqiang Lan,
		Xiao Tang,
        Ruonan Zhang,
        Bin Li,
        Qinghe Du,
        Dusit Niyato,
        and Zhu Han

\thanks{X. Lan, R. Zhang, and B. Li are with the School of Electronics and Information, Northwestern Polytechnical University, Xi'an 710072, China. (e-mail: lanxunqiang@mail.nwpu.edu.cn, rzhang@nwpu.edu.cn, libin@nwpu.edu.cn)}
\thanks{X. Tang and Q. Du are with School of Information and Communication Engineering, Xi'an Jiaotong University, Xi'an 710049, China. (e-mail: tangxiao@xjtu.edu.cn, duqinghe@mail.xjtu.edu.cn)}
\thanks{D. Niyato with the College of Computing and Data Science, Nanyang Technological University, Singapore. (email: dniyato@ntu.edu.sg)}
\thanks{Z. Han is with the Department of Electrical and Computer Engineering at the University of Houston, Houston, TX 77004 USA, and also with the Department of Computer Science and Engineering, Kyung Hee University, Seoul 446-701, South Korea. (e-mail: hanzhu22@gmail.com)}
}

\maketitle

\begin{abstract}
Blockchain plays a crucial role in ensuring the security and integrity of decentralized systems, with the proof-of-work (PoW) mechanism being fundamental for achieving distributed consensus. As PoW blockchains see broader adoption, an increasingly diverse set of miners with varying computing capabilities participate in the network. In this paper, we consider the PoW blockchain mining, where the miners are associated with resource uncertainties. To characterize the uncertainty computing resources at different mining participants, we establish an ambiguous set representing uncertainty of resource distributions. Then, the networked mining is formulated as a non-cooperative game, where distributionally robust performance is calculated for each individual miner to tackle the resource uncertainties. We prove the existence of the equilibrium of the distributionally robust mining game. To derive the equilibrium, we propose the conditional value-at-risk (CVaR)-based reinterpretation of the best response of each miner. We then solve the individual strategy with alternating optimization, which facilitates the iteration among miners towards the game equilibrium. Furthermore, we consider the case that the ambiguity of resource distribution reduces to Gaussian distribution and the case that another uncertainties vanish, and then characterize the properties of the equilibrium therein along with a distributed algorithm to achieve the equilibrium. Simulation results show that the proposed approaches effectively converge to the equilibrium, and effectively tackle the uncertainties in blockchain mining to achieve a robust performance guarantee.
\end{abstract}


\section{Introduction}
\IEEEPARstart{B}{lockchain} technology provides the fundamental framework for distributed ledgers that underpin applications such as cryptocurrencies~\cite{Bitc}, smart grids~\cite{Sg}, healthcare~\cite{Hc}, knowledge acquisition ecosystem~\cite{kae}, military systems~\cite{Ml}, vehicular networks~\cite{Vn}, and the Internet of Things (IoT)~\cite{IoT}. Its security is rooted in a combination of decentralization, cryptographic hashing, and consensus mechanisms, which together ensure data integrity, transparency, and tamper resistance. Proof of Work (PoW) consensus dominates public blockchain deployments due to its mathematically established resistance to Byzantine failures~\cite{Ml}, where network participants (miners) execute computationally intensive operations to validate transaction blocks through cryptographic puzzle solving. This process imposes exponential resource costs for malicious ledger alterations, effectively preventing double-spending attacks and ensuring transaction irreversibility. Validated blocks undergo decentralized verification before permanent blockchain integration, creating tamper-evident transaction records with cryptographic auditability~\cite{Scs}. Nevertheless, the high computation and energy demands associated with solving the puzzles pose challenges, limiting the deployment of PoW-based blockchains on general-purpose devices.

To enable the implementation of PoW-based blockchain on resource-constrained devices, prior studies have commonly estimated the processing demands of mining operations and employed auxiliary computing resources such as cloud platforms, mining pools, and edge or fog computing to relieve the local computational burden~\cite{Cloud-Net}. For instance, in~\cite{Edge} and~\cite{fog}, computationally intensive consensus tasks are offloaded to edge servers, and overall system utility is improved by optimizing miners' service demands and resource pricing. However, as blockchain technology continues to expand into a wider range of applications, the diversity and heterogeneity among participating entities have become increasingly evident.
In addition to conventional mining hardware, heterogeneous computational resources such as hosted computation, multi-source resource scheduling, and emerging embedded compute units within cloud infrastructures, vehicular platforms, and IoT systems are increasingly becoming effective sources of computational capacity that miners can leverage for PoW-based blockchain computation. These devices are often subject to dynamic workloads, leading to fluctuating and sometimes unpredictable computing availability~\cite{MultiObj}. For instance, vehicular platforms can offer surplus compute resources externally, but safety-critical task demands cause the available capacity to fluctuate with traffic, environment, and workloads.
Moreover, external factors such as thermal throttling due to overheating or voltage instability in battery-powered systems can further affect the stability of available computational resources. Therefore, it can be seen that the prevailing reliance on deterministic estimations of computing resources overlooks the inherent uncertainty in practical systems~\cite{Noun1}~\cite{Noun2}. Incorporating resource variability into the design of blockchain mechanisms is essential for developing robust and adaptive resource allocation strategies capable of operating effectively under dynamic conditions.

Unlike conventional network resource allocation schemes, mining in PoW-based blockchain networks inherently involves competitive dynamics~\cite{Com-Secure}~\cite{Bound-Net}. Therefore, optimizing resource allocation strategies must account for the impact of other mining nodes. In particular, the natural competition among miners drives them to incur higher costs to secure rewards, while each miner must balance the potential revenue from increased computational power against the accompanying expenses. This interaction forms the basis of the mining game, where miners strategically decide the amount of computing resources to invest in mining to maximize their utility. Most existing works formulate mining as a game-theoretic problem where miners optimize resource allocation and service providers set resource prices to maximize their utilities. For example, in~\cite{pg1}, the authors modeled miner-service provider interactions as a multi-leader multi-follower Stackelberg game, where miners allocate computing requests, and service providers adjust pricing competitively. In~\cite{pg2}, the authors integrated coalition formation games to optimize miner grouping and resource trading with cloud service providers.

However, existing models typically assume that miners possess perfectly known and stable computational resources, thereby neglecting the uncertainties introduced by dynamic workloads, hardware limitations, and environmental variations. Such simplifications can result in suboptimal resource allocation, imprecise pricing mechanisms, and diminished mining performance. While prior works have addressed resource uncertainty through the worst-case optimization in network settings, these approaches rely on predefined worst-case bounds that are often impractical to ascertain, given the multifaceted nature of uncertainty sources~\cite{NR}. To overcome these issues, we propose a distributionally robust game framework for PoW-based blockchain mining. By utilizing historical data to extract the statistical characteristics of resource uncertainty, our approach constructs a probabilistic description that underpins more robust mining strategies.

To the best of our knowledge, this work is among the first to model mining competition under computing resource uncertainty in blockchain systems. A primary challenge resides in analyzing equilibrium when miners' strategies are influenced by uncertain resource availability, thereby complicating conventional game-theoretic analysis. To address this, we develop a distributionally robust optimization framework for equilibrium analysis that extends deterministic game theory to incorporate uncertainty. The main contributions of this work are summarized as follows:
\begin{itemize}
\item For the proposed PoW-based blockchain network, we address the challenge of accurately obtaining the uncertainty distribution of miners' available computing resources by constructing an uncertainty model based on statistical information. The mining competition among miners is then formulated as a non-cooperative, distributionally robust game. Furthermore, we formulate the problem as a distributionally robust chance-constrained optimization and design miners' resource allocation strategies to maximize individual utility.
\item To solve the distributionally robust optimization problem in which each miner aims to maximize their expected utility under the worst-case distribution, we use a Conditional Value-at-Risk (CVaR)-based method to approximate the chance constraints as deterministic constraints. Then, a robust alternating optimization algorithm is developed to obtain the distributionally robust strategy. Finally, we design a best-response iteration algorithm for the distributionally robust Nash equilibrium.
\item We analyze the equilibrium under the special cases of the uncertain distribution reduces to Gaussian-based game and the deterministic game. Specifically, we introduce a Bernstein-type inequality (BTI)-based method to convert the chance constraints in the optimization problem into deterministic constraints. We then solve the optimization problem using an alternating optimization algorithm and obtain the game equilibrium under Gaussian distribution through best-response iterations. Furthermore, we analyze the existence and uniqueness of the game in the deterministic setting.
\item Simulation results demonstrate that, despite the inherent uncertainty in available computing resources, the proposed distributionally robust optimization framework significantly outperforms traditional mining strategies in terms of both profitability and system stability. Under scenarios of limited computational capacity, conventional strategies tend to expose miners to high risks regarding profitability. In contrast, our framework provides a reliable mechanism for resource distribution that not only mitigates these risks but also supports the long-term scalability of PoW blockchain systems.
\end{itemize}

The rest of this paper is organized as follows. In Sec.~\ref{sec:rw}, we review the related works. In Sec.~\ref{sec:sys}, we introduce the PoW-based blockchain mining model and the uncertainty model of computing resources. In Sec.~\ref{sec:Dromg}, we analyze the mining game equilibrium under an uncertain resource model, solve the distributionally robust problem using a CVaR-based method, and then introduce an alternating optimization algorithm to address the mining game. In Sec.~\ref{sec:Sc}, we present both the Gaussian-based game and the deterministic game models and analyze their respective game equilibria. Sec.~\ref{sec:sim} provides the simulation results to demonstrate the performance, and finally, Sec.~\ref{sec:con} concludes this paper.
\vspace{-1.0mm}

\section{Related Works} \label{sec:rw}
\subsection{Blockchain Mining}
Recently, there has been increasing interest in developing mining mechanisms in blockchain. In~\cite{relate1}, the authors proposed Relay-PoW, a cooperative PoW mechanism that enhances energy efficiency, resource utilization, and throughput through edge-managed mining and a Shapley-based reward strategy. In~\cite{relate9}, the authors proposed a cloud mining pool-aided blockchain architecture enabling IoT devices to offload mining and addressed pool selection via a centralized evolutionary game and a distributed reinforcement learning algorithm. In~\cite{relate10}, the authors formulated a collaborative mining network for mobile blockchain and modeled resource allocation using a double auction game and a Stackelberg game to optimize mining utility and cloud operator profits. In~\cite{a3c}, the authors employed an asynchronous advantage actor-critic deep reinforcement learning algorithm and prospect theory to optimize resource pricing and allocation for mobile blockchain mining. In~\cite{cca}, the authors proposed agent and cloud mining approaches for blockchain-enabled IoT, formulating resource allocation as a joint optimization problem and solving it using dueling deep reinforcement learning. In~\cite{tto}, the authors proposed a blockchain and mobile edge computing (MEC) integration framework with a collaborative mining process and a reputation-based PoW consensus mechanism to enhance mining efficiency and transaction throughput. In~\cite{joe}, the authors proposed a blockchain-based incentive scheme for content caching, where an edge computing server provides computing power for mining and different pricing and reward mechanisms are designed to optimize cache quality and content distribution. In~\cite{iov}, the authors proposed a blockchain-enabled internet of vehicles framework that offloads PoW tasks to MEC servers, optimizing computing efficiency.
\vspace{-3.0mm}

\subsection{Mining Games}
A lot of recent studies model the mining process in PoW-based blockchain networks as a noncooperative game, where rational miners withhold their newly found blocks with valid PoW solutions to obtain a high expected payoff. In~\cite{relate2}, the authors modeled MEC-assisted PoW mining as a stochastic Stackelberg game, where a service provider sets hash rate prices, miners independently optimize their decisions, and a hierarchical learning framework ensures stable strategies. In~\cite{relate5}, the authors formulated a multi-leader multi-follower Stackelberg game for resource management in a two-layer offloading paradigm and analyzed equilibrium under fixed and dynamic miner numbers. In~\cite{relate7}, the authors modeled mining pool selection in PoW-based blockchains as an evolutionary game, analyzing stability under different strategies and validating results through numerical simulations. In~\cite{relate8}, the authors formulated a three-stage Stackelberg Game to analyze the interactions among the blockchain protocol designer, users, and miners. In addition, some studies have explored robust blockchain systems under incomplete/imperfect information. In~\cite{ddr}, the authors modeled pricing and resource management in a blockchain-MEC system as a stochastic Stackelberg game under incomplete information and proposed hierarchical RL and deep learning algorithms to achieve stable strategies. In~\cite{lbm}, the authors modeled PoW mining in an MEC-enabled blockchain as a stochastic Stackelberg game under private information and proposed a hierarchical learning framework to achieve stable miner-service provider interactions. In~\cite{relate3}, the authors formulated a stochastic game for PoW mining in MEC-assisted blockchains under incomplete information and proposed Bayesian reinforcement learning and deep learning algorithms to dynamically adjust strategies and achieve equilibrium. In~\cite{bap}, the authors proposed a consortium blockchain-based personalized car insurance scheme that models contract-based auditing as a recursive inspection game under imperfect information to detect data fraud. While previous works have demonstrated that considering incomplete/imperfect information can enhance system robustness, there is a dearth of mining schemes for blockchain networks that comprehensively consider uncertainties in miners' computing resources.

\section{System Model} \label{sec:sys}
\subsection{Blockchain Mining Model}
\begin{figure}[t]
	\centering
	\includegraphics[width=9.0cm]{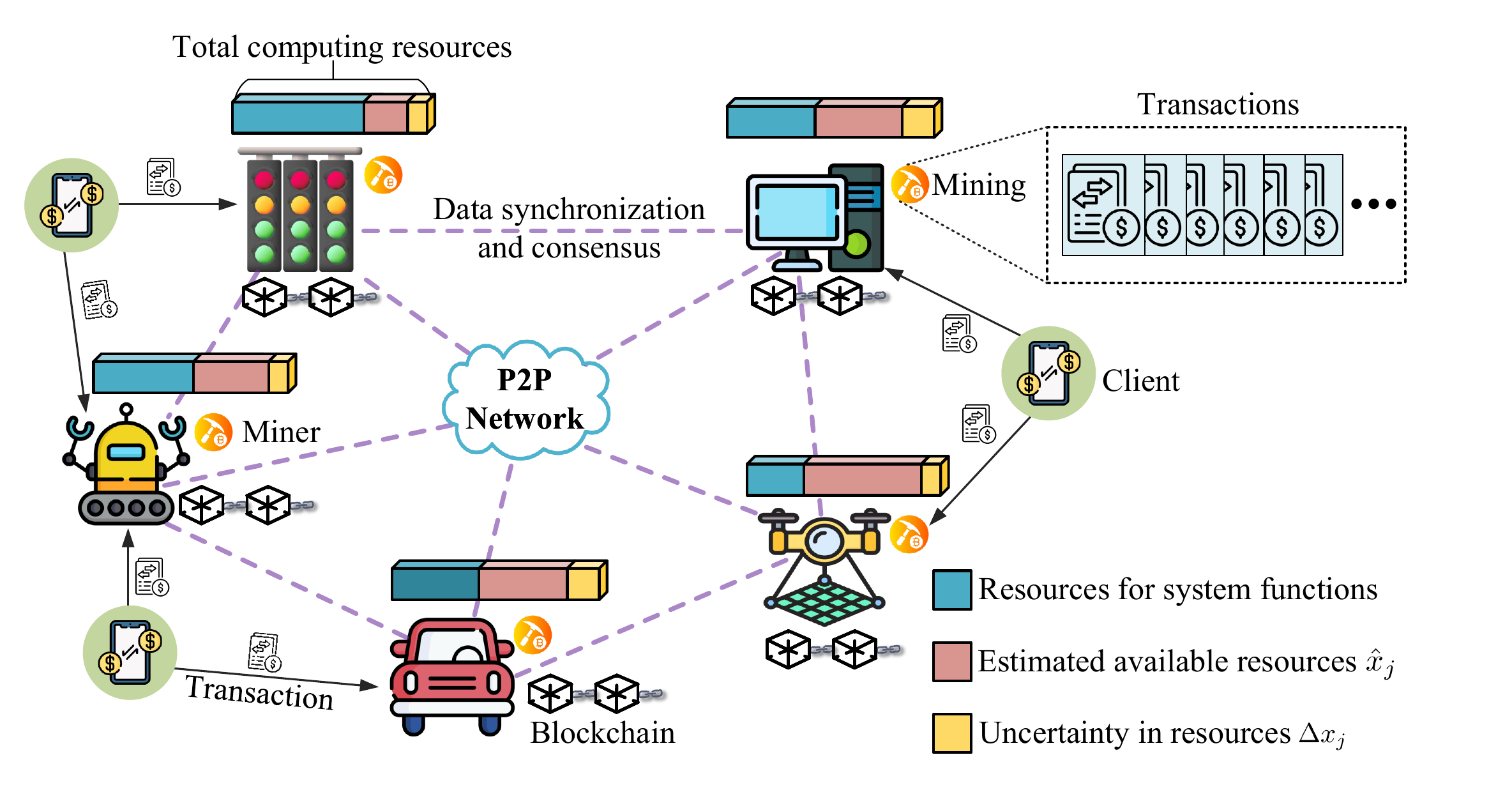}
	\vspace{-6.0mm}
	\caption{System model.}
	\vspace{-2.0mm}
	\label{fig:sys}
\end{figure}

We consider a public blockchain network under the PoW-based consensus protocol as shown in Fig.~\ref{fig:sys}. The blockchain network serves as the backbone for specific decentralized applications, providing reliable support for data storage, operational logic, transaction management, and decentralized governance in fields such as the IoT, smart transportation, smart cities, and smart healthcare. In the blockchain network, a set of $J$ nodes with a certain level of local computing power are interested in participating in the consensus process as miners and make extra profit through block mining, which is denoted as $\mathcal{J} = \left\{1,2, \dots, J\right\}$. In particular, these miners need to perform computing tasks for normal functional operations while mining. Therefore, they can only allocate the necessary hashing power for mining from their remaining available computing resources, where the available computing resources of miners are
\begin{equation}
	\bm{\mathbf{x}} = \left\{x_1, x_2, \dots , x_J \right\},\quad\forall j \in \mathcal{J},
\end{equation}
where the confidence interval for $x_j$ is $\left[x^{\min}, x^{\max}\right]$, with $x^{\min}$ and $x^{\max}$ representing the lower and upper confidence bounds on a miner’s available computing resources derived from historical data. In practice, $x^{\min}$ corresponds to the redundant resources reserved for managing dynamic or complex situations, while $x^{\max}$ denotes the maximum available resources determined by the miner’s total capacity.

In the blockchain network, the generation of a new block consists of two stages: mining and consensus. In the mining stage, miners compete against each other in order to be the first one to solve the PoW puzzle and obtain the consequent mining reward from the speed game accordingly. Therefore, the probability that miner-$j$ successfully mining a block is proportional to its relative computing power (hashing power) $h_j$, which is the ratio of the hashing power allocated by miner-$j$ to the overall hashing power of all miners. It is defined as follows:
\begin{equation}
	h_{j}\left(\alpha_j, \bm{\alpha}_{-j}\right) = \frac{\alpha_j x_j}{\sum_{k \in \mathcal{J}}{\alpha_k x_k}},
\end{equation}
where $\alpha_j\in (\tau_0,1]$ is the mining investment coefficient of miner-$j$, $\tau_0>0$, $\bm{\alpha}_{-j}$ represents the mining investment coefficient of all other miners except miner-$j$, and $\alpha_j x_j$ is the computing resources allocated to mining by miner-$j$, $\forall j\in\mathcal{J}$ and thus we have $h_j>0$, $\sum_{k \in \mathcal{J}}{h_k}=1$. $\tau_0$ is the minimum threshold of $\alpha_j$, aimed at preventing miners from strategically exiting, thereby avoiding system instability and ensuring the continuous participation of each miner in maintaining the normal operation of the blockchain system. In addition, without a centralized authority, a consensus mechanism is required to validate new blocks. When a miner solves the puzzle, they propagate the block to the network. Other miners then verify it and reach consensus on appending it to the blockchain. It is worth noting that although the present discussion focuses on PoW, the proportional selection framework remains applicable to proof-of-stake (PoS) and delegated proof-of-stake (DPoS) systems when the mining capability is interpreted as effective stake or delegated voting weight.

The first miner to successfully mine a block that reaches consensus earns the reward. The reward consists of two parts, one is a fixed reward $\Phi$, and the other is a commission reward $\rho \Psi$, totally $\left(\Phi + \rho \Psi \right)$. Here, $\rho$ represents the reward for the unit transaction, and $\Psi$ denotes the number of mining transactions. Meanwhile, the resources utilized for mining also incur corresponding operational costs, including CPU time and energy costs, with the cost per unit of resource at miner-$j$ denoted as $c_j$. Therefore, the expected utility of miner-$j$ is given as
\begin{equation} \label{eq:util}
	U_{j}\left(\alpha_j, \bm{\alpha}_{-j}\right) = \left(\Phi + \rho \Psi \right)\frac{\alpha_j x_j}{\sum_{k \in \mathcal{J}}{\alpha_k x_k}} - c_j \alpha_j x_j,
\end{equation}
where $c_j \alpha_j x_j$ is the corresponding mining resource costs of miner-$j$, $\forall j\in\mathcal{J}$.

\subsection{Uncertainty Model}
In practice, due to various uncertainty factors, there are usually errors between the estimated available computing resources of miners in blockchain networks and their actual available computing resources. The uncertainty is primarily reflected in several aspects: First, miners' concurrent tasks may require unpredictable data processing, causing volatile resource consumption. Second, dynamic node participation in mining pools alters task demands and available resources. Third, processor throttling under high load reduces computing power, and fourth, unstable power supplies further contribute to resource fluctuations. Given these factors, miners in the blockchain network often struggle to accurately predict the actual available computing resources when planning their resources. This can result in either over-allocation or under-allocation of resources, which in turn impacts their probability of successful mining and profitability. We also note that the PoS and DPoS systems exhibit analogous variability due to dynamic staking, delegation changes, and slashing events, which similarly affect a participant's effective ability to compete in block production.

In our proposed model, we consider the uncertainty of available computing resources to make a robust design. In the resource uncertainty case, the actual available computing resources of miners can be expressed as
\begin{equation}
	x_j = \hat x_j + \Delta x_j,\quad\forall j \in \mathcal{J},
\end{equation}
where $ \hat x_j $ is the estimated available computing resources of miner-$j$ based on historical observations; $ \Delta x_j $ is the uncertainty of available computing resources, which follows a specific random distribution.

Due to the combined influence of various random factors with different attributes, it is often challenging to obtain the exact distribution information. In contrast, we can easily obtain the first and second-order statistical parameters of the uncertainty distribution of available computing resources based on historical data. Motivated by this, we define a ambiguity set $\mathcal{P}_j$ of all distributions of $\Delta x_j$ with the same means and variances, as follows
\begin{equation} \label{eq:fuzzy}
	\mathcal{P}_j = \left\{\mathbb{P}_j: \mathbb{E}_{\mathbb{P}_j}\left(\Delta x_j\right)=\mu_j,\mathbb{E}_{\mathbb{P}_j}\left[\left(\Delta x_j-\mu_j\right)^2\right]=\sigma^2_j\right\},
\end{equation}
where $\mathbb{E}_{\mathbb{P}_j}(\cdot)$ denotes expectation with respect to $\mathbb{P}_j$, $\mu_j$ and $\sigma^2_j$ are the means and variances of $\Delta x_j$ under distribution $\mathbb{P}_j$, and $\mathbb{P}_j$ can be arbitrary distribution as long as it meets the mean and variance requirements in~(\ref{eq:fuzzy}), i.e. $\mathbb{P}_j\in\mathcal{P}_j\left(\mu_j,\sigma^2_j\right)$, $\forall j \in \mathcal{J}$. In practice, miners continuously track their effective hash rate, and the resulting fluctuations form a locally measurable stochastic process. Owing to its ergodicity, each node can estimate the mean and variance of the uncertainty through temporal averaging of past observations.

\section{Distributionally Robust Mining Game} \label{sec:Dromg}
In this section, we model the mining competition in the PoW-based blockchain network as a non-cooperative, distributionally robust game, where each miner aims to optimize their utilities. We then analyze the Nash equilibrium of the distributionally robust game.
\subsection{Game Model Formulation}
In the competition of mining, miners are only concerned with their own expected utility, based on which their investment in mining is determined. Furthermore, considering the uncertainty in computing resources, we incorporate the uncertainty model~(\ref{eq:fuzzy}) into~(\ref{eq:util}) and formulate the following distributionally robust optimization problem for each miner: maximizing the individual expected utility under the worst-case distribution, i.e.,
\begin{IEEEeqnarray}{cl}
	\IEEEyesnumber\label{eq:p_mf} \IEEEyessubnumber*
	\max_{\alpha_j} \quad & \inf_{\mathbb{P}_j\in\mathcal{P}_j}~\mathbb{E}_{\Delta x_j \sim\mathbb{P}_j} U_j\left(\alpha_j, \bm{\alpha}_{-j}, \Delta \bm{x}\right)  \\
	\rm{s.t.} \quad & \tau_0 \leq \alpha_j \leq 1, \quad \forall j\in\mathcal{J}, \label{eq:alpha_c}
\end{IEEEeqnarray}
where $\inf_{\mathbb{P}_j \in \mathcal{P}_j}$ denotes the lower bound of the probability under the probability distribution $\mathbb{P}_j$, $\mathbb{E}_{\mathbb{P}_j}$ denotes the expectation induced by the probability distribution $\mathbb{P}_j$, $U_j\left(\alpha_j, \bm{\alpha}_{-j}, \Delta \bm{x}\right)$ is the utility function of miner-$j$ and $\Delta \bm{x}=\left[\Delta x_1,\Delta x_2, \cdots, \Delta x_J\right]$ is the set of uncertainties in the available computing resources of all miners. The constraint in~(\ref{eq:alpha_c}) gives the constraint for each miner's mining investment coefficient.

Considering that the distributionally robust problem in~(\ref{eq:p_mf}) needs to be solved at each miner, there naturally follows the non-cooperative game formulation. Therefore, we introduce the distributionally robust game defined as follows.

\textit{Definition 1:} The distributionally robust mining game $\mathcal{G}$ for the PoW-based blockchain network is given by
\begin{equation} \label{eq:dro_game}
	\mathcal{G} = \left\{\mathcal{J}, \left\{\mathcal{A}_j\right\}_{j \in \mathcal{J}}, \left\{\inf_{\mathbb{P}_j\in\mathcal{P}_j}~\mathbb{E}_{\mathbb{P}_j} U_j\left(\alpha_j, \bm{\alpha}_{-j}, \Delta \bm{x}\right)\right\}_{j \in \mathcal{J}}\right\},
\end{equation}
which is comprised of,
\begin{itemize}
	\item $\mathcal{J}$: The set of players, i.e., the miners;
	\item $\mathcal{A}_j$: The strategy space at miner-$j$, i.e., the set of the mining investment coefficient $\alpha_{j}$ that satisfies~(\ref{eq:alpha_c});
	\item $\inf_{\mathbb{P}_j\in\mathcal{P}_j}~\mathbb{E}_{\mathbb{P}_j} U_j\left(\alpha_j, \bm{\alpha}_{-j}, \Delta \bm{x}\right)$: The individual utility function that quantifies the preference of miner-$j$ under the worst-case distribution.
\end{itemize}

We can see that in the distributionally robust game, miners lack complete information about the true probability distribution of potential uncertainties but must formulate their strategies under such uncertainty. Specifically, each miner leverages partial information to construct a set of candidate distributions and optimizes its decision based on the worst-case distribution rather, thereby mitigating the risks associated with distributional ambiguity. Moreover, when the underlying data of these decision problems contain uncertain parameters, the resulting distributionally robust Nash equilibrium problem becomes significantly more challenging than a standard Nash equilibrium problem.

\subsection{Equilibrium Analysis}
The solution to the distributionally robust game is characterized by the Nash equilibrium~\cite{Gameth}, which corresponds to the steady states of the miners’ mining competition in the game. In our considered game $\mathcal{G}$, we analyze the distributionally robust Nash equilibrium, which can be formally defined as follows.

\textit{Definition 2 (Distributionally Robust Nash Equilibrium):} For the game $\mathcal{G}$, a tuple $\bm{\alpha}^*=\left(\alpha^*_1, \alpha^*_2, \dots , \alpha^*_J\right)$ is called a Nash equilibrium of the distributionally robust Nash equilibrium if
\begin{equation}
	\alpha^*_j\in \arg \max_{\alpha_j \in \mathcal{A}_j}\inf_{\mathbb{P}_j \in \mathcal{P}_j}\mathbb{E}_{\mathbb{P}_j}\left[U_j\left(\alpha_j, \bm{\alpha}^*_{-j}, \Delta \bm{x}\right)\right],~j \in \mathcal{J}.
\end{equation}

Next, we analyze the existence of the distributionally robust Nash equilibrium, which represents the optimal computing power allocation strategy that enables miners to maintain reasonable utility under the worst-case distribution, mitigating losses due to probability misestimation. This equilibrium enhances stability compared to the traditional Nash equilibrium, making miners' decisions less susceptible to fluctuations caused by uncertainty.

\begin{theorem}\label{thm1}
	A Nash equilibrium exists in the distributionally robust game $\mathcal{G}$.
\end{theorem} 

\begin{IEEEproof} Based on the equilibrium theory of distributionally robust game~\cite[Theorem~2.1]{DROCC}, a Nash equilibrium exists in the distributionally robust game ${\mathcal{G}}$ when the following assertions hold for all miners.
\begin{enumerate}[label=(\arabic*), leftmargin=3em]
	\itshape
	\item $U_j(\cdot)$ is a continuous function and $U_j(\cdot, \bm{\alpha}_{-j}, \Delta \bm{x})$ is a concave function of $\alpha_j$ for every $(\bm{\alpha}_{-j}, \Delta \bm{x})$;
	\item $\mathcal{A}=\prod_{j\in\mathcal{J}}\mathcal{A}_j$ is a compact set;
	\item $\mathbb{E}_{\mathbb{P}_j} [U_j(\alpha_j, \bm{\alpha}_{-j}, \Delta \bm{x})]$ is finite valued for any $\alpha \in \mathcal{A}$ and $\mathbb{P}_j \in \mathcal{P}_j$;
	\item $\mathcal{P}_j$ is weakly compact.
\end{enumerate}

First, for the conditions in~(1) and~(2), the strategy space $\mathcal{A}_j$ for each miner is defined as the interval $[\tau_0,1]$, which is a nonempty, convex, and compact subset of the Euclidean space. It is evident that the overall strategy space $\mathcal{A}$ is also compact. Considering the utility function $U_j$ as given by~(\ref{eq:util}), it is evident that $U_j$ is continuous over the interval $[\tau_0,1]$. Subsequently, to prove the concavity of $U_j$, we derive the first-order and second-order derivatives of~(\ref{eq:util}) with respect to $\alpha_j$, as detailed below.
\begin{equation} \label{eq:u1}
	\frac{\partial U_{j}}{\partial\alpha_{j}}=x_{j}(\Phi+\rho\Psi)\frac{\sum_{l \neq j}\alpha_{l}x_{l}}{(\sum_{k\in\mathcal{J}}\alpha_{k}x_{k})^{2}}-c_{j}x_{j},
\end{equation}
\begin{equation} \label{eq:u2}
	\frac{\partial^{2} U_{j}}{\partial\alpha_{j}^{2}}=-2x^2_{j}(\Phi + \rho\Psi)\frac{\sum_{l \neq j}\alpha_{l}x_{l}}{(\sum_{k\in\mathcal{J}}\alpha_{k}x_{k})^{3}}.
\end{equation}
Since $\frac{\sum_{l \neq j}\alpha_{l}x_{l}}{(\sum_{k\in\mathcal{J}}\alpha_{k}x_{k})^{2}} > 0$ and $-2x^2_{j}\frac{\sum_{l \neq j}\alpha_{l}x_{l}}{(\sum_{k\in\mathcal{J}}\alpha_{k}x_{k})^{3}}<0$, the second order derivative of $U_j$ with respect to $\alpha_j$ is strictly negative for any $(\bm{\alpha}_{-j}, \Delta \bm{x})$. Therefore, $U_j$ is concave in $\alpha_j$.

Second, as $U_j(\alpha, \Delta \bm{x})$ is concave with respect to $\alpha_j$ regardless of the uncertainties, its expectation operator $\mathbb{E}_{\mathbb{P}_j} [U_j(\alpha, \Delta \bm{x})]$ is also concave of $\alpha_j$. Additionally, since $\alpha_j\in [\tau_0,1]$, this ensures the boundedness of $\mathbb{E}_{\mathbb{P}_j} [U_j(\alpha, \Delta \bm{x})]$ in the direction of $\alpha_j$. The boundedness of $\bm{\alpha}_{-j}$ further restricts the variation range across other dimensions. Consequently, because $[\tau_0,1]^{J}$ constitutes a compact set, $\mathbb{E}_{\mathbb{P}_j} [U_j(\alpha, \Delta \bm{x})]$ is bounded over this domain, i.e. condition~(3) holds. Finally, for condition~(4), by applying the Prokhorov's theorem and considering the boundedness of the mean and variance, it follows that the uncertainty set $\mathcal{P}_j$ is compact in the weak topology. Therefore, there exists a Nash equilibrium in the proposed distributionally robust game. The best-response strategy can be achieved through sequential iterative updates by miners.
\end{IEEEproof}

The distributionally robust equilibrium model is mathematically complex, requiring novel approaches beyond existing numerical methods to identify the Nash equilibrium of the distributionally robust game. To address this, we develop an alternating iterative method to obtain the equilibrium solution. Specifically, we first consider the optimization problem of miner-1, assuming the strategies of all other miners, denoted as $\bm{\alpha}_{-1}$, remain fixed. By solving optimization problem of miner-1 in~(\ref{eq:p_mf}), the following optimal solution (The best response) is obtained: $\alpha^*_1\in \arg \max_{\alpha_1 \in \mathcal{A}_1}\inf_{\mathbb{P}_1 \in \mathcal{P}_1}\mathbb{E}_{\mathbb{P}_1}\left[U_1\left(\alpha_1, \bm{\alpha}_{-1}, \Delta \bm{x}\right)\right]$. Subsequently, miner-1 updates its strategy to $\alpha^*_1$, and the same procedure is sequentially applied to optimize and update the strategies of the remaining miners. This iterative process continues until all miners' strategies converge, yielding the Nash equilibrium of the distributionally robust game~\cite{DROCC}.

\subsection{CVaR-Based Reinterpretation}
Based on the above analysis of the distributionally robust equilibrium, we next focus on solving the distributionally robust optimization problem for miner-$j$, given that the strategies of all other miners $\bm{\alpha}_{-j}$ remain fixed. Due to the non-concave objective function, solving the distributionally robust problem in~(\ref{eq:p_mf}) is particularly challenging. In this subsection, we introduce a CVaR-based method to address this problem.

For the distributionally robust problem in~(\ref{eq:p_mf}), It is obvious that the uncertainty in available computing resources presents a significant challenge in designing a system that consistently satisfies each miner's expected utility requirements. In practice, it is reasonable to introduce the minimum utility threshold $U ^ {\ min} _j $ and adopt a distributionally robust design under uncertainty to ensure that $U_j\left(\alpha_j, \bm{\alpha}_{-j}, \Delta \bm{x}\right) \geq U^{\min}_j$ with a certain probability. Therefore, we reformulate the problem in~(\ref{eq:p_mf}) as a distributionally robust chance-constrained problem, i.e.,
\begin{IEEEeqnarray}{cl}
	\IEEEyesnumber\label{eq:pp0} \IEEEyessubnumber*
	\max_{\alpha_j, U^{\min}_j} \quad & U^{\min}_j  \\ \nonumber
	\rm{s.t.} \quad & \inf_{\mathbb{P}_j \in \mathcal{P}_j}\Pr\nolimits_{\mathbb{P}_j}\left[\frac{\left(\Phi + \rho \Psi \right)\alpha_j x_j}{\sum_{k\in\mathcal{J}}{\alpha_k x_k}} - c_j \alpha_j x_j \geq U^{\min}_j\right] \\ 
	& \geq 1-\epsilon, \quad \forall j\in\mathcal{J}, \label{eq:Pr} \\ \nonumber
	&(\text{\ref{eq:alpha_c}}),
\end{IEEEeqnarray}
where $U^{\min}_j$ is the minimum utility threshold of miner-$j$. The constraint in~(\ref{eq:Pr}) is the distributionally robust chance constraint, ensuring that under uncertain available computing resources, the utility of miner-$j$ exceeds the minimum utility threshold $U_j^{\min}$ with a probability of at least $1-\epsilon$, where $\epsilon$ is the tolerance probability.

For the distributionally robust chance constraint in~(\ref{eq:Pr}), the CVaR-based method can be applied as the tightest convex approximation~\cite{cvar2}. Specifically, when the constraint function exhibits concavity or quadratic behavior with respect to the random variable, the distributionally robust version of chance constraints is equivalent to the worst-case CVaR constraint~\cite{cvar3}, which is described by the follows.

As for the utility chance constraint in~(\ref{eq:Pr}), we have the following continuous loss function:
\begin{equation}
	\begin{split}
		L(x_j)= &c_j \alpha_j^2 x_j^2+\bigg[U^{\min}_{j} - \left(\Phi + \rho \Psi \right) \bigg.\\
		&\bigg.+ c_j\sum_{l \neq j}{\alpha_l x_l}\bigg]\alpha_j x_j + U^{\min}_{j}\sum_{l \neq j}{\alpha_l x_l},
	\end{split}
\end{equation}
which is a quadratic function with respect to $x_j$. Then, the distributionally robust chance constraint is equivalent to the worst-case constraint, given by
\begin{equation}\label{eq:cvar_lem}
	\begin{split}
		\inf_{\mathbb{P}_j \in \mathcal{P}_j}& \Pr\nolimits_{\mathbb{P}_j}\left\{L(x_j) \leq 0\right\} \geq 1 - \epsilon\\
		&\Leftrightarrow \sup_{\mathbb{P}_j \in \mathcal{P}_j} \mathbb{P}_j\text{--}\text{CVaR}_\epsilon \left\{L(x_j)\right\} \leq 0,
	\end{split}
\end{equation}
where $\mathbb{P}_j\text{--}\text{CVaR}_\epsilon \left\{L(x_j)\right\}$ is denoted as the CVaR of $L(x_j)$ at threshold $\epsilon$ with respect to $\mathbb{P}_j$, defined as
\begin{equation}\label{eq:cvar_lem1}
	\mathbb{P}_j\text{--}\text{CVaR}_\epsilon \left\{L(x_j)\right\} = \inf_{\beta_j\in \mathbb{R}}\left\{\beta_j + \frac{1}{\epsilon}\mathbb{E}_{\mathbb{P}_j}\left[\left(L(x_j)-\beta_j\right)^{+}\right] \right\}.
\end{equation}
Moreover, $(\cdot)^+ = \max \left\{0, \cdot\right\}$, and $\beta_j \in \mathbb{R}$ is an auxiliary variable introduced by CVaR.

From~(\ref{eq:cvar_lem}), CVaR can quantify the conditional expectation of loss exceeding the $ (1-\epsilon) $-quantile of the loss distribution of the random variable $L(x_j)$ by constructing convex approximations of distributionally robust chance constraints. The worst-case CVaR can be converted into a group of semi-definite programs (SDPs)~\cite{cvar3}. Therefore, the chance constraint in~(\ref{eq:Pr}) can be transformed as the following CVaR constraint:
\begin{equation}\label{eq:cvar_all}
	\left\{
	\begin{aligned}
		&\beta_j + \frac{1}{\epsilon}\text{Tr}\left(\bm{\Omega}_j \bm{ M}_j\right) \leq 0,~\beta_j \in \mathbb{R},\\
		&\bm{M}_j \in \mathbb{S}^{2},~\bm{M}_j\succcurlyeq 0,\\
		&\bm{M}_j-
		\begin{bmatrix}
			c_j\alpha_j^2&\frac{1}{2}\alpha_j(U^{\min}_j + B_j)\\ 
			\frac{1}{2}\alpha_j(U^{\min}_j + B_j)&U^{\min}_{j}\sum_{l \neq j}{\alpha_l x_l}-\beta_j
		\end{bmatrix}\succcurlyeq 0,
	\end{aligned}
	\right.
\end{equation}
where $\bm{ M}_j$ and $\beta_j$ are the auxiliary variables and $\bm{M}_j \succeq 0$ indicates that $\bm{M}_j$ is a positive-semidefinite matrix, $\mathbb{S}^2$ denotes the space of $2$-dimensional symmetric matrix, and $\bm{\Omega}_j$ is a matrix defined as:
\begin{equation}
	\bm{\Omega}_j=\begin{bmatrix}
		\sigma^{2}_j+\bar{\mu}_j^2&\bar{\mu}_j\\
		\bar{\mu}_j&1
	\end{bmatrix}, 
\end{equation}
\begin{equation}
	\bar{\mu}_j=\hat x_j + \mu_j,
\end{equation}
where $\bar{\mu}_j \in \mathbb{R}$ and $\sigma^2_j \in \mathbb{R}$ denote the mean and variance of random variable $x_j$, respectively.

Therefore, the original distributionally robust chance-constrained problem in~(\ref{eq:pp0}) can be reformulated as
\begin{IEEEeqnarray}{cl}
	\IEEEyesnumber\label{eq:p_cvar}
	\max_{\alpha_j, U^{\min}_j, \bm{M}_j, \beta_j} \quad & U^{\min}_j  \\ \nonumber
	\rm{s.t.} \quad &(\text{\ref{eq:alpha_c}}), (\text{\ref{eq:cvar_all}}).
\end{IEEEeqnarray}
Now the chance constraint has been reformulated in a tractable deterministic form by the CVaR-based method. However, the utility constraints still remain non-convex.  In the subsequent step, we develop an alternating optimization (AO) algorithm to solve the problem in~(\ref{eq:p_cvar}).

\subsection{AO-Based Algorithm}
In this subsection, we design an alternating optimization method to address the optimization problem. The coupling between variables $\alpha_j$ and $U_j^{min}$ results in the inclusion of non-linear matrix inequality constraint in~(\ref{eq:p_cvar}), rendering a non-convex optimization problem. Therefore, it is necessary to decouple the optimization variables $\alpha_j$ and $U_j^{min}$ by decomposing the optimization problem into two subproblems, optimizing the mining strategy $\alpha_j$ and the utility threshold $U_j^{min}$ separately. Then, obtaining an approximate solution to the optimization problem by alternating framework.

Based on the above analysis, we decompose the problem in~(\ref{eq:p_cvar}) into the following two subproblems. First, we consider~(\ref{eq:p_cvar}) with fixed the mining investment coefficient $\alpha_j$, the problem can be rewrite as:
\begin{IEEEeqnarray}{cl}
	\IEEEyesnumber\label{eq:p-c1} \IEEEyessubnumber*
	\max_{\begin{subarray}{l}
			U^{\min}_j, \\
			\bm{M}_j, \beta_j
	\end{subarray}} \quad & U^{\min}_j  \\
	\rm{s.t.} &\beta_j + \frac{1}{\epsilon}\text{Tr}\left(\bm{\Omega}_j \bm{ M}_j\right) \leq 0, \\
	&\beta_j \in \mathbb{R}, \bm{M}_j \in \mathbb{S}^{2}, \bm{M}_j\succcurlyeq 0,\\
	&\bm{M}_j \succcurlyeq\!
	\begin{bmatrix}
		c_j \alpha_j^2&\frac{\alpha_j}{2}(U^{\min}_j + B_j)\\ 
		\frac{\alpha_j}{2}(U^{\min}_j + B_j)&U^{\min}_{j}\sum\limits_{l \neq j}{\alpha_l x_l}-\beta_j
	\end{bmatrix}\!.\quad\quad
\end{IEEEeqnarray}
Since the problem in~(\ref{eq:p-c1}) is convex, it can be efficiently solved using a convex optimization solver, e.g. CVX. Then, we fix the optimal minimum expected individual utility $U_j^{\min}$, two auxiliary variables $\bm{M}_j$ and $\beta_j$. The problem in~(\ref{eq:p_cvar}) can be reformulated as
\begin{IEEEeqnarray}{cl}
	\IEEEyesnumber\label{eq:p-c2} \IEEEyessubnumber*
	\max_{\alpha_j} \quad & U^{\min}_j  \\
	\text{s.t.} &\bm{M}_j \succcurlyeq
	\begin{bmatrix}
		c_j \alpha_j^2&\frac{\alpha_j}{2}(U^{\min}_j + B_j)\\ 
		\frac{\alpha_j}{2}(U^{\min}_j + B_j)&U^{\min}_{j}\sum\limits_{l \neq j}{\alpha_l x_l}-\beta_j
	\end{bmatrix},\quad\quad \label{eq:psub2_con} \\ \nonumber
	&(\text{\ref{eq:alpha_c}}).
\end{IEEEeqnarray}
However, constraint~(\ref{eq:psub2_con}) is still a non-linear matrix inequality constraint, which needs further handling. Hence, we introduce an auxiliary variable $t^c_j$ with the aim of transforming constraint~(\ref{eq:psub2_con}) into a linear matrix inequality (LMI), thereby converting the problem in~(\ref{eq:p-c2}) into the following convex optimization problem as
\begin{IEEEeqnarray}{cl}
	\IEEEyesnumber\label{eq:p-c3} \IEEEyessubnumber*
	\max_{\alpha_j, t^c_j} \quad & U^{\min}_j  \\
	\text{s.t.} &\bm{M}_j \succcurlyeq
	\begin{bmatrix}
		t^c_j&\frac{\alpha_j}{2}(U^{\min}_j + B_j)\\ 
		\frac{\alpha_j}{2}(U^{\min}_j + B_j)&U^{\min}_{j}\sum\limits_{l \neq j}{\alpha_l x_l}-\beta_j
	\end{bmatrix},\quad\quad \label{eq:psub3_con} \\
	&c_j \alpha_j^2 \leq t^c_j,\\ \nonumber
	&(\text{\ref{eq:alpha_c}}).
\end{IEEEeqnarray}
Now, an approximate solution to problem in~(\ref{eq:p_cvar}) can be obtained by alternately optimizing problems in~(\ref{eq:p-c1}) and~(\ref{eq:p-c3}) to update $\alpha_j$ and $U^{\min}_j$, respectively.

Based on the CVaR-based method above, we employ the best-response iteration algorithm to obtain the distributionally robust equilibrium. In this process, each miner iteratively adjusts its strategy in response to the current strategies of other miners, which guides the iterations toward a best response consistent stationary point or a local Nash equilibrium. Specifically, given the resource allocation strategy $\bm{\alpha}_{-j}$ of all other miners, miner-$j$ determines its best response by solving~(\ref{eq:p_cvar}). Each miner sequentially updates its strategy with the CVaR-based method, optimizing its objective while accounting for uncertainty in available resources. This iterative best-response adaptation ensures that the collective strategy set gradually stabilizes, ultimately leading to the distributionally robust Nash equilibrium, detailed procedure is shown in Alg.~\ref{alg:1}.

\section{Special-Case Investigation} \label{sec:Sc}
As the previously formulated distributionally robust mining game relies on strong assumptions about the ambiguity set, increasing its complexity, we explore two practical scenarios: one where computing resource uncertainty follows a Gaussian distribution and another with deterministic resources. These scenarios represent typical simplifications of dynamic and stable network environments, respectively.

\subsection{Gaussian Mining Game}
According to the central limit theorem, as the number of uncertain factors increase, the uncertainty of available computing resources of miners can be modeled as follow the Gaussian distribution~\cite{CLT}.

\begin{algorithm}
	\caption{Best-Response Iteration Algorithm for Distributionally Robust Nash Equilibrium}
	\label{alg:1}
	\SetKwData{In}{\textbf{in}}\SetKwData{To}{to}
	\DontPrintSemicolon
	\SetAlgoLined
	\KwIn {Select initial input $\alpha_j\in [\tau_0,1]$, $U_j^{\min}$ according to~(\ref{eq:util}), $j \in \mathcal{J}$, tolerance probability $\epsilon$, precision threshold $\kappa$, $t \leftarrow 1$.}
	\Repeat{$\lVert \sum\limits_{j \in \mathcal{J}}\left(\alpha_j[t]-\alpha_j\right) + \sum\limits_{j \in \mathcal{J}}\left(U^{\min}_j[t]-U^{\min}_j\right)\rVert \leq \kappa$.}{
		Let $U_j^{\min}[t] = U_j^{\min}$ and $\alpha_j[t] = \alpha_j$, $j \in \mathcal{J}$;\\
		{\For{each $j \in \mathcal{J}$}{
				Solve problem in~(\ref{eq:p_cvar}) by alternately optimizing problems~(\ref{eq:p-c1}) and~(\ref{eq:p-c3}), obtain the optimal $U_j^{\min*}$ and $\alpha_j^*$;
		}}
		$t = t+1$
	}
	\KwOut {The optimal mining strategy $\bm{\alpha}^{*}$.}
\end{algorithm}

\subsubsection{Game Formulation and Equilibrium}
In the case of a Gaussian distribution, since the distribution type is known, each miner's distributionally robust optimization problem in~(\ref{eq:p_mf}) can be reformulated as:
\begin{IEEEeqnarray}{cl}
	\IEEEyesnumber\label{eq:p_bmf} \IEEEyessubnumber*
	\max_{\alpha_j} \quad & \mathbb{E}_{\mathbb{P}_j} U_j\left(\alpha_j, \bm{\alpha}_{-j}, \Delta \bm{x}\right)  \\
	\rm{s.t.} \quad & \tau_0 \leq \alpha_j \leq 1, \quad \forall j\in\mathcal{J}, \label{eq:alpha_b}
\end{IEEEeqnarray}
where $\mathbb{P}_j=\mathcal{N}\left(\mu_j,\sigma^2_j\right)$. Then, the corresponding Gaussian mining game is defined as a non-cooperative game in which each miner maximizes its expected utility under Gaussian-distributed uncertainty in computing resources, as follows.

\textit{Definition 3:} The Gaussian mining game $\widetilde{\mathcal{G}}$ for the PoW-based blockchain network is given by
\begin{equation} \label{eq:gm_game}
	\widetilde{\mathcal{G}} = \left\{\mathcal{J}, \left\{\mathcal{A}_j\right\}_{j \in \mathcal{J}}, \left\{\mathbb{E}_{\mathbb{P}_j} U_j\left(\alpha_j, \bm{\alpha}_{-j}, \Delta \bm{x}\right)\right\}_{j \in \mathcal{J}}\right\},
\end{equation}
which is comprised of,
\begin{itemize}
	\item $\mathcal{J}$: The set of players, i.e., the miners;
	\item $\mathcal{A}_j$: The strategy space at miner-$j$, i.e., the set of the mining investment coefficient $\alpha_{j}$ that satisfies~(\ref{eq:alpha_b});
	\item $\mathbb{E}_{\mathbb{P}_j} U_j\left(\alpha_j, \bm{\alpha}_{-j}, \Delta \bm{x}\right)$: The individual utility function that quantifies the preference of miner-$j$ under the Gaussian distribution.
\end{itemize}

Since the Gaussian mining game is a special case of the distributionally robust mining game, the existence of an equilibrium in the Gaussian mining game can be readily verified, so we omit the details here for space limitation.

\subsubsection{BTI-Based Analysis and Algorithm}
Due to the mathematical complexity of the equilibrium model in the Gaussian mining game, an alternating iterative approach is needed to obtain the equilibrium solution. Specifically, the game equilibrium is achieved by sequentially updating each miner's best response, where the best response of miner-$j$ is obtained by solving the optimization problem in~(\ref{eq:p_bmf}) while keeping the strategies of the other miners, $\bm{\alpha}_{-j}$, fixed. To address the Gaussian optimization problem in~(\ref{eq:p_bmf}), we introduce the minimum utility threshold $U ^ {\ min} _j $ and reformulate the problem as the following chance-constrained problem:
\begin{IEEEeqnarray}{cl}
	\IEEEyesnumber\label{eq:ppb0} \IEEEyessubnumber*
	\max_{\alpha_j, U^{\min}_j} \quad & U^{\min}_j  \\ \nonumber
	\rm{s.t.} \quad & \Pr\nolimits_{\mathbb{P}_j}\left[\frac{\left(\Phi + \rho \Psi \right)\alpha_j x_j}{\sum_{k\in\mathcal{J}}{\alpha_k x_k}} - c_j \alpha_j x_j \geq U^{\min}_j\right] \\ 
	& \geq 1-\epsilon, \quad \forall j\in\mathcal{J}, \label{eq:Prb} \\ \nonumber
	&(\text{\ref{eq:alpha_b}}).
\end{IEEEeqnarray}
Then, we use the Bernstein-type inequality approach to provide a safe approximation for the problem in~(\ref{eq:ppb0}), which bounds the probability that a sum of random variables deviates from its mean~\cite{BTI1}. The Bernstein-type inequality allows us to convert the probabilistic constraints into deterministic forms, which is summarized as follows.

Considering the scenario of the Gaussian distribution, we define the uncertainty variable of the available resources as $\Delta x_j \sim \mathcal{N}(\mu_j,\sigma^2_j)$. Let $\Delta x_j=\mu_j+\sigma_j e_j$, where $e_j \sim \mathcal{N}(0,1)$. The utility chance constraint in~(\ref{eq:Prb}) can be reformulated as
\begin{equation} \label{eq:bti_pr}
	\Pr\nolimits_{\mathbb{P}} \left\{ f(e_j) \geq 0 \right\} \geq 1 - \epsilon,~\forall j \in \mathcal{J},
\end{equation}
where
\begin{equation} \label{eq:bti_f}
	f(e_j)= A_j e_j^2 +2b_j e_j + D_j,
\end{equation}
\begin{equation} \label{eq:bti_A}
	A_j= -c_j \alpha_j^2 \sigma_j^2,
\end{equation}
\begin{equation} \label{eq:bti_b}
	b_j= -\sigma_j \left[c_j \alpha_j^2 \left(\hat{x}_j+\mu_j\right) + \frac{1}{2}\left(U_j^{\min}+B_j\right)\alpha_j\right],
\end{equation}
\begin{equation} \label{eq:bti_B}
	B_j= -(\Phi+\rho \Psi) + c_j\sum_{l \neq j}\alpha_l x_l,
\end{equation}
\begin{equation} \label{eq:bti_D}
	\begin{aligned}
		D_j=&-\bigg[c_j \alpha_j^2 \left(\hat{x}_j+\mu_j\right)^2 + \left(U_j^{\min}+B_j\right)\alpha_j\left(\hat{x}_j+\mu_j\right) \bigg.\\
		&\bigg. +U_j^{\min}\sum_{l \neq j}\alpha_l x_l\bigg].
	\end{aligned}
\end{equation}
According to the Bernstein-type inequality~\cite{BTI1}, let $\delta=-\ln{\epsilon}$, then the chance constraint form in~(\ref{eq:Prb}) can be conservatively rewritten as the following deterministic form,
\begin{equation} \label{eq:bti_deter}
	A_j - \sqrt{-2 \ln(\epsilon)}\sqrt{|A_j|^{2}+2|b_j|^{2}} + \ln(\epsilon) s^{+}A_j + D_j \geq 0.
\end{equation}
However, the constraint in~(\ref{eq:bti_deter}) is still not a linear constraint. By introducing two auxiliary variables $\upsilon_j$, $\omega_j$, we transform the constraint in~(\ref{eq:bti_deter}) into the following system of the LMI and Second-Order Cone (SOC) constraints:
\begin{subnumcases} {\label{weqn}}
A_j - \sqrt{-2 \ln(\epsilon)}\,\upsilon_j + \ln(\epsilon)\omega_j + D_j \ge 0, \label{eq:bti_v1v2}\\
\left\|
    \begin{bmatrix}
        A_j\\
        \sqrt{2}b_j
    \end{bmatrix}
\right\| \le \upsilon_j, \label{eq:bti_v1}\\
\omega_j + A_j\ge 0,\quad \omega_j\ge 0. \label{eq:bti_v2}
\end{subnumcases}
Therefore, the chance-constrained problem in~(\ref{eq:ppb0}) is reformulated as a tractable optimization problem, given by
\begin{IEEEeqnarray}{cl}
	\IEEEyesnumber\label{eq:p_bti}
	\max_{\alpha_j, U^{\min}_j, \upsilon_j, \omega_j} \quad & U^{\min}_j  \\ \nonumber
	\rm{s.t.} \quad & (\text{\ref{eq:alpha_c}}), (\text{\ref{eq:bti_v1v2}}), (\text{\ref{eq:bti_v1}}), (\text{\ref{eq:bti_v2}}),
\end{IEEEeqnarray}
where the constraints in~(\ref{eq:bti_v1v2})$\sim$(\ref{eq:bti_v2}) are the transformation of the utility chance constraint in~(\ref{eq:Prb}).

Although the chance constraint has been reformulated into deterministic form, the optimization problem remains non-convex due to the multiplication of $\alpha_j$ and $U_j^{\min}$ in the constraints. Therefore, it is necessary to decouple the optimization variables $\alpha_j$ and $U_j^{min}$ using an alternating iterative optimization method to optimize the mining strategy $\alpha_j$ and the utility threshold $U_j^{min}$ separately.

Then, we decompose the problem in~(\ref{eq:p_bti}) into the following two subproblems. First, fix the variable $\alpha_j$ and solve the problem in~(\ref{eq:p_bti}), which can be rewritten as a convex optimization problem as follows:
\begin{IEEEeqnarray}{cl}
	\IEEEyesnumber\label{eq:p_b1}
	\max_{U^{\min}_j, \upsilon_j, \omega_j} \quad & U^{\min}_j  \\ \nonumber
	\rm{s.t.} \quad & (\text{\ref{eq:bti_v1v2}}), (\text{\ref{eq:bti_v1}}), (\text{\ref{eq:bti_v2}}).
\end{IEEEeqnarray}
Since the problem in~(\ref{eq:p_b1}) is convex, it can be directly solved. Then, we fix the optimal minimum expected individual utility $U_j^{\min}$, two auxiliary variables $v_j$ and $w_j$. The problem in~(\ref{eq:p_bti}) can be reformulated as
\begin{IEEEeqnarray}{cl}
	\IEEEyesnumber\label{eq:p_b2}
	\max_{\alpha_j} \quad & U^{\min}_j  \\  \nonumber
	\rm{s.t.} \quad & (\text{\ref{eq:alpha_b}}), (\text{\ref{eq:bti_v1v2}}), (\text{\ref{eq:bti_v1}}), (\text{\ref{eq:bti_v2}}).
\end{IEEEeqnarray}
However, $\alpha_j^2$ introduces non-convexity in constraints~(\ref{eq:bti_v1v2})$\sim$(\ref{eq:bti_v2}), necessitating further handling. Hence, we introduce an auxiliary variable $t^b_j$ to convert the problem in~(\ref{eq:p_b2}) into the following convex optimization problem.
\begin{IEEEeqnarray}{cl}
	\IEEEyesnumber\label{eq:p_b3}  \IEEEyessubnumber*
	\max_{\alpha_j, t^b_j} \quad & U^{\min}_j  \\
	\rm{s.t.} \quad & \bar{A}_j - \sqrt{-2 \ln(\epsilon)}\upsilon_j + \ln(\epsilon)\omega_j + \bar{D}_j \geq 0, \\
	& \left \| \begin{bmatrix}
		\bar{A}_j\\
		\sqrt{2}\cdot\bar{b}_j
	\end{bmatrix} \right \| \leq \upsilon_j,~\omega_j + \bar{A}_j\geq 0,\\ \nonumber
	& -c_j\alpha_j^2\geq t^b_j,~(\text{\ref{eq:alpha_b}}),
\end{IEEEeqnarray}
where
\begin{equation}
	\left\{
	\begin{aligned}
		&\bar{A}_j= t^b_j \sigma_j^2,\\
		&\bar{b}_j= \sigma_j \left[t^b_j \left(\hat{x}_j+\mu_j\right) - \frac{1}{2}\left(U_j^{\min}+B_j\right)\alpha_j\right],\\
		&\begin{aligned}
			\bar{D}_j=&t^b_j \left(\hat{x}_j+\mu_j\right)^2 - \left(U_j^{\min}+B_j\right)\alpha_j\left(\hat{x}_j+\mu_j\right)\\
			&-U_j^{\min}\sum_{l \neq j}\alpha_l x_l.
		\end{aligned}
	\end{aligned}
	\right.
\end{equation}
Here $t^b_j$ is an auxiliary variable bounding $-c_j\alpha_j^2$. By solving the problem in~(\ref{eq:p_b3}), we can obtain the optimal mining
investment coefficient $\alpha_{j}^*$. Finally, by alternately optimizing problems in~(\ref{eq:p_b1}) and~(\ref{eq:p_b3}) to update $\alpha_j$ and $U^{\min}_j$, respectively, an approximate solution to the problem in~(\ref{eq:p_bti}) can be obtained.

Based on the optimization problems formulated using the BTI-based method described above, we propose an alternating optimization algorithm to obtain the best response of game $\widetilde{\mathcal{G}}$. The best-response algorithm can be broadly described as follows: given the computing resources investment strategy $\bm{\alpha}_{-j}$ of other miners, miner-$j$ updates its strategy $\alpha_j$ by solving either~(\ref{eq:p_bti}). By iteratively updating each miner's strategy, the algorithm gradually converges to the game equilibrium.

\begin{table}[h]\centering
	\caption{Simulation Parameters}
	\label{table}
	\renewcommand{\arraystretch}{1.5}
	\begin{tabular}{c|c|c}
		\specialrule{1.5pt}{0pt}{0pt}
		\hline
		\textbf{Parameter} & \textbf{Description} & \textbf{Value} \\
		\hline
		$\Phi$ & Fixed reward  & 5000 \\
		$\rho$ & Reward for unit transactions & 10 \\
		$\Psi$  & The number of mining transactions & 300 \\
		$c_j$  & Cost per unit of resource at miner-$j$ & 60 \\
		$\epsilon$  & Tolerable probability & 0.1 \\
		$\kappa$  & Precision threshold & $10^{-6}$ \\
		$\tau_0$  & Minimum mining investment coefficient & 0.5 \\
		$x^{\min}$  & Redundant minimum computing resources & 10 \\
		$x^{\max}$ & Maximum available computing resources & 100 \\
		\hline        
	\end{tabular}
\end{table}
\subsection{Deterministic Game}
Under deterministic computing resources, miners strategically adjust their investment in mining computational power to maximize their own utility, reflecting their selfish nature. Mathematically, the optimization problem at miner $j$ can be expressed as
\begin{IEEEeqnarray}{cl}
	\IEEEyesnumber\label{eq:p0} \IEEEyessubnumber*
	\max_{\alpha_j} \quad & U_j\left(\alpha_j, \bm{\alpha}_{-j}\right)  \\
	\rm{s.t.} \quad	&  \tau_0 \leq \alpha_j \leq 1, \quad \forall j\in\mathcal{J}. \label{eq:alpha_d}
\end{IEEEeqnarray}
Then, under deterministic computing resources, all miners design their strategies based on the deterministic available resources to maximize their expected utility, which transforms the utility optimization problem in~(\ref{eq:p0}) to a deterministic optimization problem. Accordingly, the mining in the blockchain network is modeled as the following non-cooperative game.

\textit{Definition 4:} The non-cooperative mining competition game $\overline{\mathcal{G}}$ for the blockchain networks is given by
\begin{equation} \label{eq:non_game}
	\overline{\mathcal{G}} = \left\{\mathcal{J},\left\{\mathcal{A}_j\right\}_{j \in \mathcal{J}},\left\{U_j\left(\alpha_j, \bm{\alpha}_{-j}\right)\right\}_{j \in \mathcal{J}}\right\},
\end{equation}
which is comprised of,
\begin{itemize}
	\item $\mathcal{J}$: The set of players, i.e., the miners;
	\item $\mathcal{A}_j$: The strategy space at miner-$j$, i.e., the set of mining investment coefficient $\alpha_j$ that satisfies~(\ref{eq:alpha_d});
	\item $U_j\left(\alpha_j, \bm{\alpha}_{-j}\right)$: The individual utility function quantifies the preference of miner-$j$ across various strategy profiles.
\end{itemize}

\textit{Definition 5 (Nash Equilibrium):} For the non-cooperative mining competition game $\overline{\mathcal{G}}$, the Nash equilibrium is denoted by $\bm{\alpha}^*=\left(\alpha^*_1, \alpha^*_2, \dots , \alpha^*_J\right)$, which is the strategy profile that no miner can further improve its utility by unilaterally deviating from the current strategy. Mathematically, the following inequality holds at all the miner-$j \in \mathcal{J}$,
\begin{equation} \label{eq:ne}
	U_j\left(\alpha^*_j, \bm{\alpha}^*_{-j}\right) \geq U_j\left(\alpha_j, \bm{\alpha}^*_{-j}\right),~\forall \alpha_j \in \mathcal{A}_j.
\end{equation}

Next, we will prove the existence and uniqueness of the Nash equilibrium for game $\overline{\mathcal{G}}$. Based on equation~(\ref{eq:u2}), we have demonstrated that $U_j$ is concave with respect to $\alpha_{j}$, thereby ensuring the existence of the Nash equilibrium. Then, the proof of equilibrium uniqueness and the fact that the best-response function is a standard function (satisfying positivity, monotonicity, and scalability) follows a similar approach to that in~\cite{fog}. Let $\alpha^{*}_{j}$ denote the mining investment coefficient of miner-$j$ that satisfies the condition $\frac{\partial U_j}{\partial \alpha_j}=0$. Then, we can get the best response function of miner-$j$ as
\begin{equation}
	\alpha^{*}_{j}=\mathcal{F} (\alpha_{j})=\left\{
	\begin{aligned}
		&\tau_0,&\frac{\zeta_j-\sum\limits_{l\neq j}\alpha_{l}x_{l} }{x_{j}}<\tau_0,\\
		&\frac{\zeta_j-\sum\limits_{l\neq j}\alpha_{l}x_{l} }{x_{j}},~&\tau_0\leq \frac{\zeta_j-\sum\limits_{l\neq j}\alpha_{l}x_{l} }{x_{j}}\leq 1,\\
		&1,&\frac{\zeta_j-\sum\limits_{l\neq j}\alpha_{l}x_{l} }{x_{j}}\geq 1,
	\end{aligned}
	\right.
\end{equation}
where $\zeta_j=\sqrt{\frac{\left(\Phi + \rho \Psi\right)\sum_{l\neq j}\alpha_{l}x_{l}}{c_j}}$. It is apparent that the function $\mathcal{F}(\alpha_{j})$ is positive, monotonic, and scalable, which is a standard function~\cite{stand_funtc}. Therefore, there is a unique Nash Equilibrium for game $\overline{\mathcal{G}}$. Moreover, to solve the unique Nash equilibrium of game $\overline{\mathcal{G}}$, we first set~(\ref{eq:u1}) to 0, yielding the following expression:
\begin{equation} \label{eq:U1}
	\frac{\sum_{k\in\mathcal{J}}\alpha_{k}x_k - \alpha_{j}x_j}{\Big(\sum_{k\in\mathcal{J}}\alpha_{k}x_k\Big)^2}=\frac{c_j}{\left(\Phi + \rho \Psi\right)}.
\end{equation}
Then, by summing both sides of~(\ref{eq:U1}) over different $j$, we can derive as follows
\begin{equation} \label{eq:U1-1}
	\frac{J-1}{\sum_{k\in\mathcal{J}}\alpha_{k}x_k}=\frac{\sum_{j\in\mathcal{J}}c_j}{\left(\Phi + \rho \Psi\right)}.
\end{equation}
Finally, by substituting~(\ref{eq:U1-1}) into~(\ref{eq:U1}), the Nash Equilibrium is obtained as follows
\begin{equation} \label{eq:UNE}
	\alpha_j^{*}=\frac{J-1}{x_j \sum_{j\in\mathcal{J}}\frac{c_j}{\Phi + \rho \Psi}}+\bigg(\frac{J-1}{\sum_{j\in\mathcal{J}}\frac{c_j}{\Phi + \rho \Psi}}\bigg)^2 \frac{c_j}{x_j \left(\Phi + \rho \Psi\right)}.
\end{equation}
\vspace{-6.0mm}

\section{Simulation Results} \label{sec:sim}
In this section, we present the convergence performance of our proposed robust alternating optimization algorithm for the mining game. Additionally, we perform numerical simulations to evaluate the effectiveness of the proposed method under various configurations and conditions. For our considered PoW-based blockchain networks, we investigate two configurations regarding homogeneous and heterogeneous device performance, corresponding to different practical scenarios. In the simulation, there are 5 miners in the blockchain system. In the homogeneous case, each miner's estimated available computing resources $\hat{x}_j$ is set to 55. In the heterogeneous case, each miner's estimated available computing resources $\hat{x}_j$ are randomly generated following a uniform distribution $U(30,60)$. The initial mining investment coefficient is set as $\alpha_j=0.35$. The main simulation parameters, summarized in Table~\ref{table}, are used as defaults unless otherwise noted.

\subsection{Convergence}
\begin{figure}[t]
	\centering
	\includegraphics[width=7.8cm]{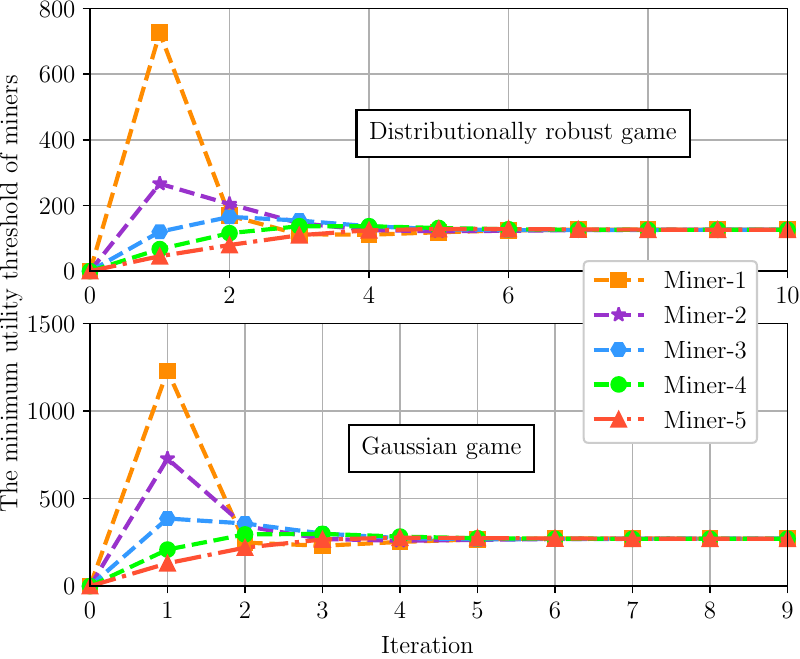}
	\vspace{-2.0mm}
	\caption{Convergence of best-response iterations in homo cases.}
	\vspace{-2.0mm}
	\label{fig:conv_ho}
\end{figure}

\begin{figure}[t]
	\centering
	\includegraphics[width=7.8cm]{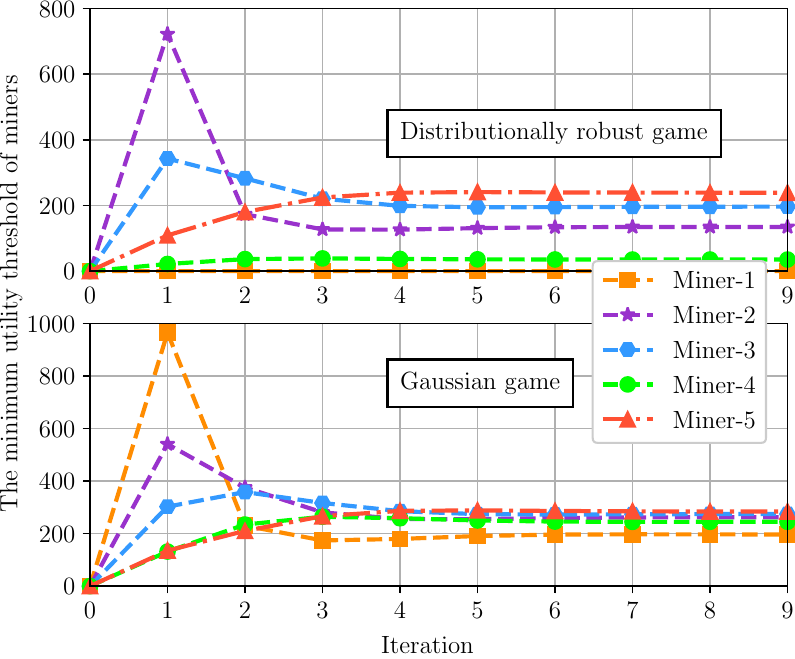}
	\vspace{-2.0mm}
	\caption{Convergence of best-response iterations in hetero cases.}
	\vspace{-5.5mm}
	\label{fig:conv_he}
\end{figure}

We first demonstrate the convergence of the proposed robust alternating optimization algorithm in Figs.~\ref{fig:conv_ho} and~\ref{fig:conv_he}, which the minimum utility threshold of miners is shown under both homogeneous and heterogeneous settings. As shown in the figures, the proposed algorithm eventually converges, and the distribution of available computing resources among miners significantly impacts the convergence behavior. In the homogeneous case, where all miners have the same available computing resources, the convergence is faster and more stable, with miner utilities reaching similar values after a few iterations. In contrast, in the heterogeneous case, where miners have varying computing power, the convergence exhibits greater fluctuations, and the utility values across miners are less synchronized due to differences in resource allocation strategies. That is because miners with higher computing power dominate resource allocation in the heterogeneous case, resulting in diverging utility values where some miners achieve significantly higher rewards than others. This highlights the impact of resource disparities in decentralized mining.
\vspace{-4.0mm}
\begin{figure}[t]
	\centering
	\includegraphics[width=7.8cm]{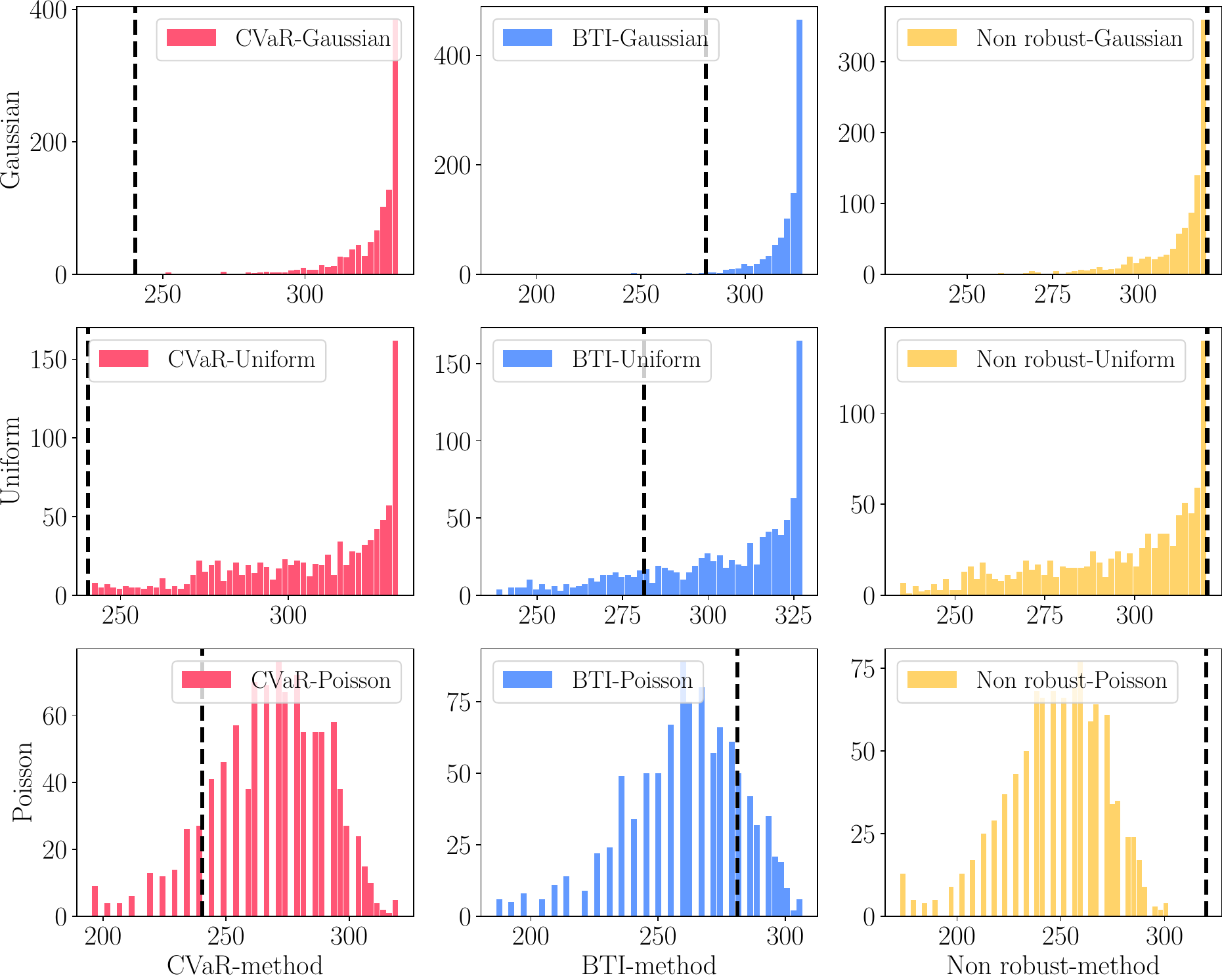}
	\caption{Histogram of miners' utilities under homogeneous case with different distributions.}
	\label{fig:hist_ho}
\end{figure}
\begin{figure}[t]
	\centering
	\includegraphics[width=7.8cm]{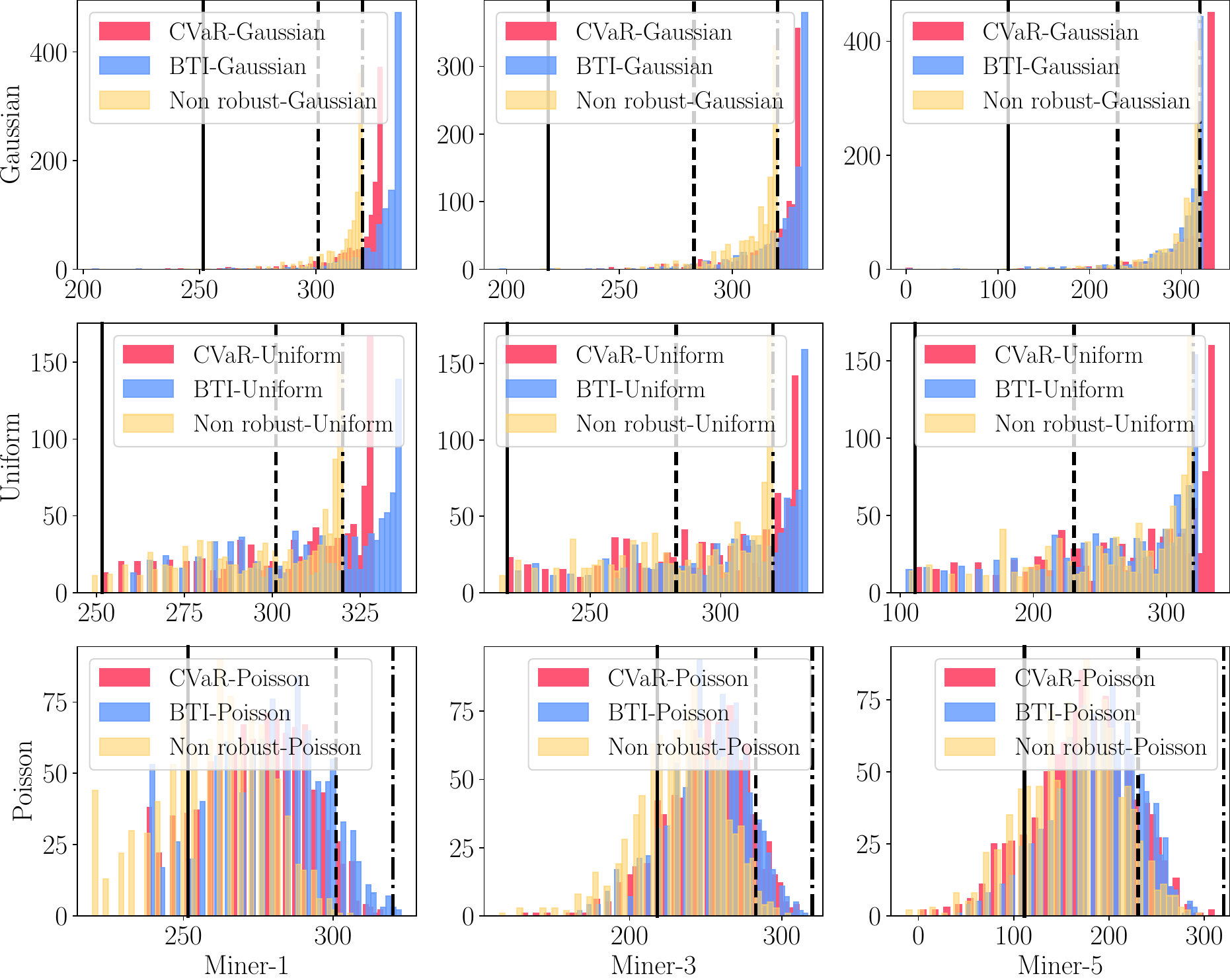}
	\caption{Histogram of miners' utilities under heterogeneous case with different distributions.}
	\vspace{-5.5mm}
	\label{fig:hist_he}
\end{figure}

\subsection{Performance Comparison}
Fig.~\ref{fig:hist_ho} demonstrates the robustness of the proposed methods by evaluating the utility of one miner under the homogeneous case, considering various uncertainty distributions of available computing resources. Since it is unrealistic to simulate all distributions under $\mathbb{P}_j\left(\mu_j, \sigma^2_j\right)$, we primarily focus on three typical distributions, namely, Gaussian, Uniform, and Poisson. In this simulation, we set the estimated available computing resources for all miners as 55, with $\mu_j=0$ and $\sigma_j=10$ of the uncertainty distribution, and the tolerance probability $\epsilon=0.1$. Specifically, we first obtain the robust optimization equilibrium solutions using the CVaR-based and BTI-based methods, while the non robust optimization equilibrium solution is obtained from~(\ref{eq:UNE}) (where $\Delta x_j = 0$). Then, for each of the three distributions, we randomly generate $10^3$ samples of $\Delta x_j$ and use these along with the equilibrium solutions to calculate the utility of miner. We plot the empirical evaluation of the utility distribution histogram in Fig.~\ref{fig:hist_ho}, where the x-axis represents the utility and the y-axis represents the number of items that meet the utility. The black dashed line represents the optimal minimum utility threshold $U^{\min}_j$. The three subplots from left to right in Fig.~\ref{fig:hist_ho} display the histograms of miner utilities under the proposed CVaR-based, BTI-based, and non robust methods. Since the miners' utilities are identical in the homogeneous scenario, only one miner's utility is presented for each distribution case here. We can observe that the empirical distribution of the proposed CVaR-based method satisfies the tolerance probability constraint, meaning that the proportion of cases where $U_j>U_j^{\min}$ exceeds $(1-\epsilon)$. In contrast, the BTI-based method, to varying degrees, violates the tolerance probability constraint. Specifically, the BTI-based method meets the tolerance probability requirement under the Gaussian distribution. This is expected, as Bernstein-type inequality is designed to accommodate Gaussian uncertainty. However, for the other two distributions, there is a violation of the probability threshold in the BTI-based design. For the non robust scheme, the utility of miner is always below the optimal utility across all tested distributions.

Fig.~\ref{fig:hist_he} demonstrates the robustness of the proposed methods in the heterogeneous case, considering three uncertainty distributions for available computing resources. For the heterogeneous case, the estimated available computing resources of miners are randomly generated according to a uniform distribution $U(30,60)$. Without loss of generality, we present the empirical utility evaluation for three of the five miners, where each column subplot represents the robustness comparison for the same miner across the different methods under different distributions. Each row subplot compares the robustness of different miners under the same distribution in the heterogeneous case. The solid and dashed lines represent the optimal minimum utility threshold $U_j^{\min}$ for the CVaR-based and BTI-based methods, respectively, while the dashdot line represents the optimal utility of the miner for the non-robust method. We observe that the CVaR-based method has the most stable utility distribution, with all miners' utilities exceeding the optimal minimum utility threshold, showing its strong robustness. In contrast, although the optimal minimum utility threshold of the BTI-based method is greater than the CVaR-based method, it satisfies the tolerance probability constraint only under the Gaussian distribution. For other distributions, such as Uniform and Poisson, the BTI-based method violates the tolerance probability constraint, resulting in miners having a significant amount of utility below the expected utility. The non-robust method performs the worst, with its utility distribution consistently skewed lower and all miners' utilities falling below the expected utility, demonstrating its vulnerability under uncertain conditions.
\begin{figure}[t]
	\centering
	\includegraphics[width=7.8cm]{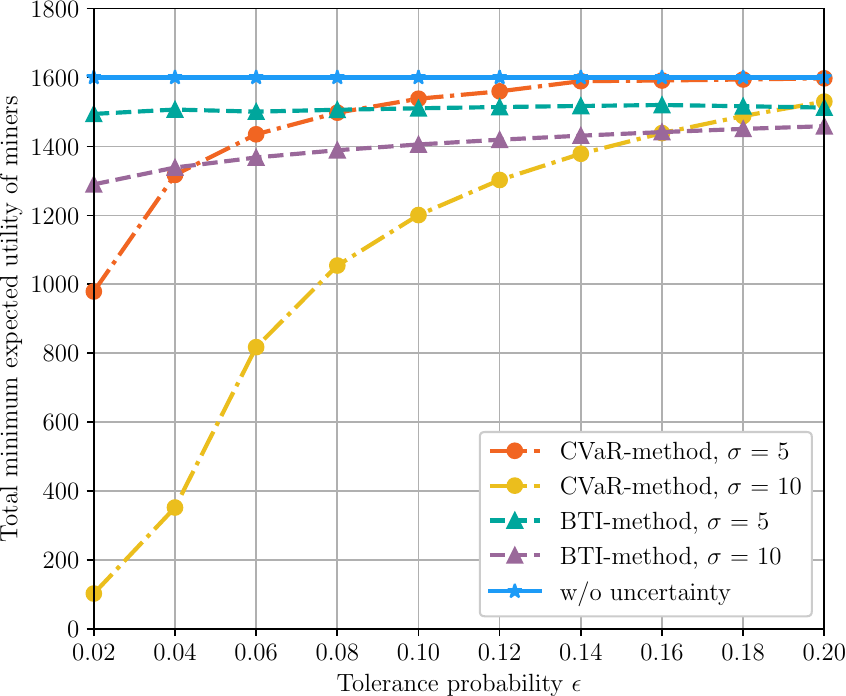}
	\caption{Total minimum expected utility of miners $\sum_{j\in\mathcal{J}}U_j^{\min}$ versus tolerance probability $\epsilon$.}
	\vspace{-4.0mm}
	\label{fig:epsilon}
\end{figure}

Fig.~\ref{fig:epsilon} illustrates the effects of threshold probability on the performance of the mining utility under different optimization methods. The simulation results show that as the tolerance probability $\epsilon$ increases, both CVaR-based and BTI-based methods increase in total minimum expected individual utility $\sum_{j\in\mathcal{J}}U_j^{\min}$, as a higher threshold allows for more aggressive strategies with a higher tolerance for deviation from the minimum expected utility. In contrast, the without uncertainty optimization method, which does not account for the uncertainty in available computing resources, maintains a constant optimal total minimum utility regardless of the threshold probability. Notably, CVaR shows lower total utility than BTI when $\epsilon$ is small, due to its more conservative approach considering the worst-case scenarios across various distributions. As $\epsilon$ grows, the performance gap narrows, and both methods converge to the almost same utility as the without uncertainty method. The impact of uncertainty is also evident, with a higher standard deviation leading to lower utility, highlighting the need for robust methods like CVaR and BTI in uncertain environments.
\begin{figure}[t]
	\centering
	\includegraphics[width=7.8cm]{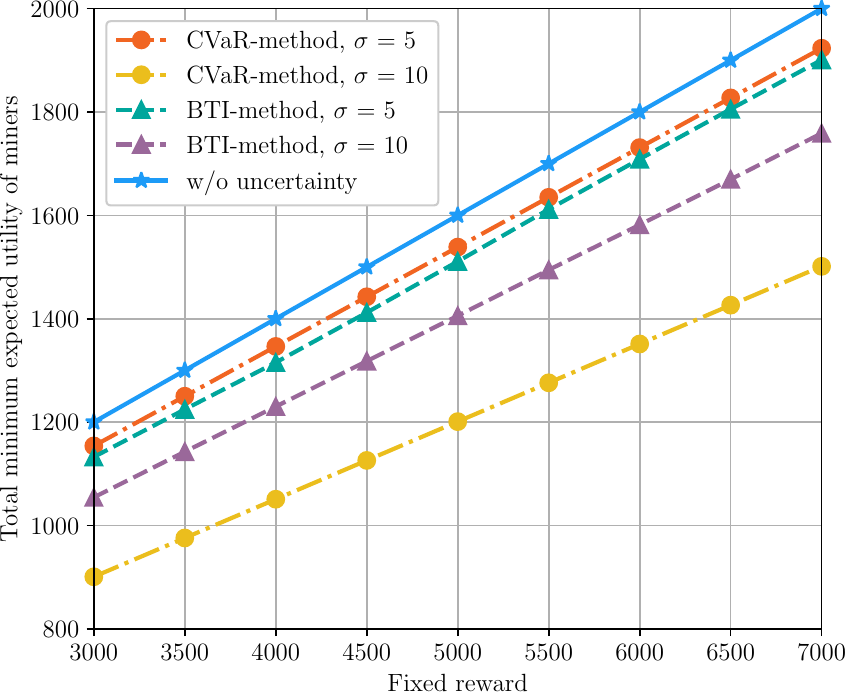}
	\caption{Total minimum expected utility of miners $\sum_{j\in\mathcal{J}}U_j^{\min}$ versus fixed reward.}
	\label{fig:reward}
\end{figure}
\begin{figure}[t]
	\centering
	\includegraphics[width=7.8cm]{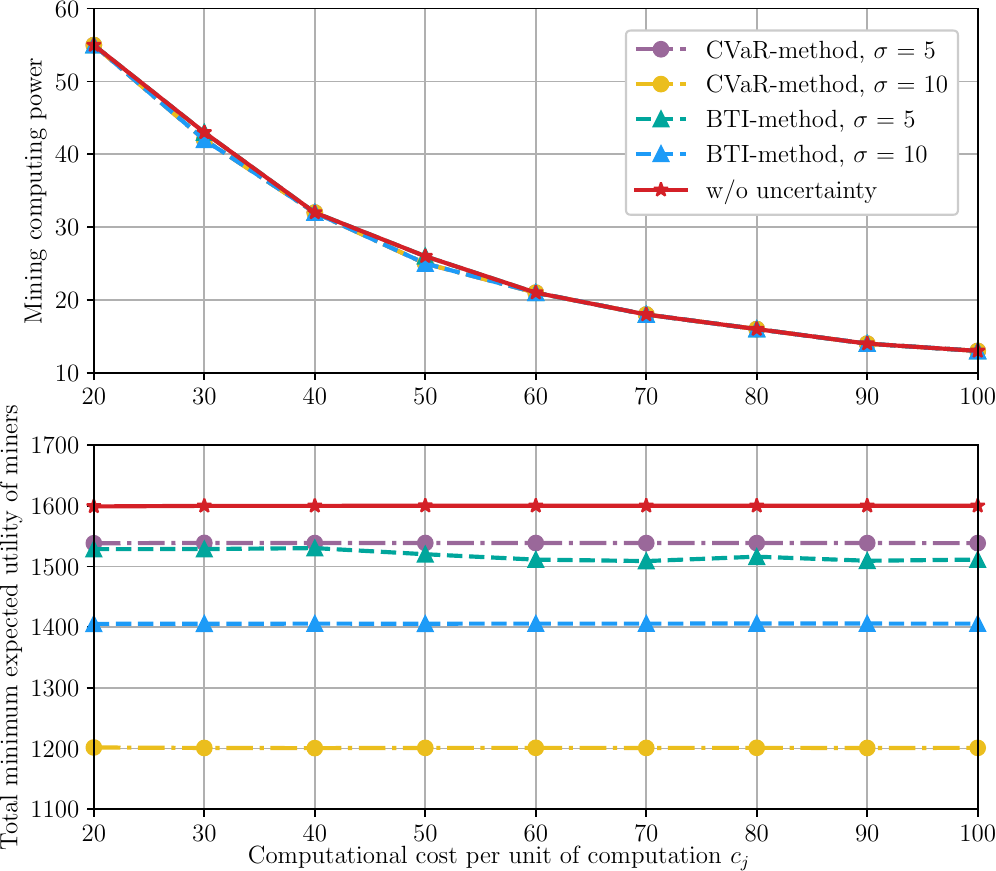}
	\caption{Total minimum expected utility of miners versus computational cost.}
	\vspace{-4.0mm}
	\label{fig:price}
\end{figure}

Fig.~\ref{fig:reward} shows the impact of different fixed rewards on the minimum expected utility of miners under different optimization methods, with a tolerance probability $\epsilon = 0.1$. As expected, as the fixed reward increases, the total minimum expected utilities of miners obtained by all methods also increase. However, excessively high fixed rewards are not reasonable, as they encourage miners to continually increase their computing resources, leading to a waste of energy and resources. Therefore, matching the appropriate reward with the resources invested is important. In addition, under lower levels of uncertainty ($\sigma = 5$), the CVaR-based method outperforms the BTI-based method and the without uncertainty method. In this case, the CVaR-based method provides more consistent and higher utility for miners, highlighting its effectiveness in handling smaller uncertainties in resource availability. Conversely, at higher levels of uncertainty, the situation is reversed. Similar to Fig.~\ref{fig:epsilon}, the without uncertainty optimization method, which does not account for uncertainty, consistently achieves the highest optimization utility.

\begin{figure}[t]
	\centering
	\includegraphics[width=7.8cm]{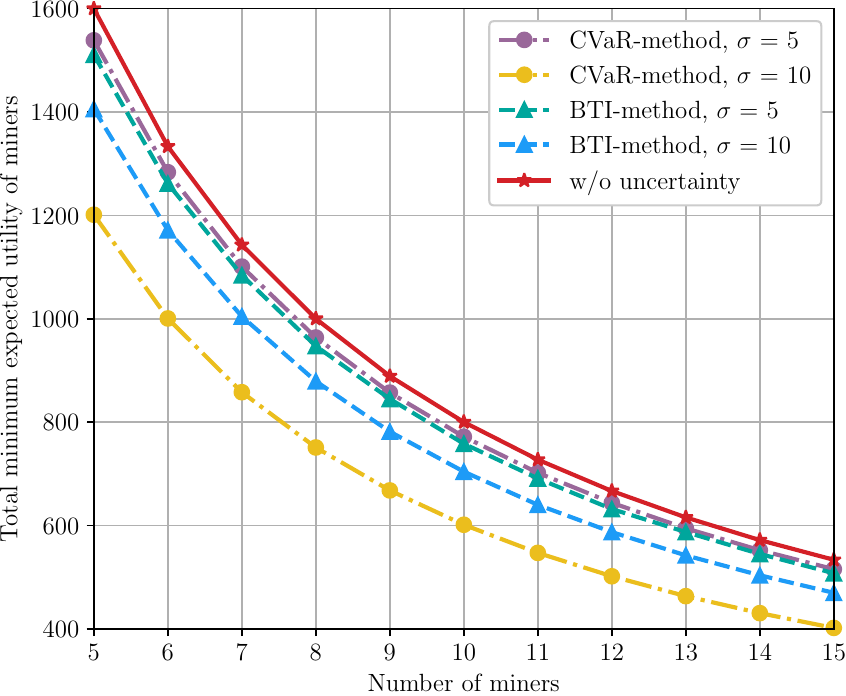}
	\caption{Performance comparison with different methods versus the number of miners under homogeneous case.}
	\vspace{-4.5mm}
	\label{fig:num_ho}
\end{figure}

In Fig.~\ref{fig:price}, we investigate the impact of varying unit computational costs on the performance of the proposed blockchain system in the homogeneous case. As shown the bottom subplot, the total utility of miners remains almost unchanged with the increase in unit computational costs. This is because, in a blockchain network with fixed overall rewards, an increase in unit computation costs raises miners' mining costs, thereby reducing their utilities. This leads to a decrease in miners' enthusiasm for mining, resulting in a reduction in their total computational contribution in order to minimize overall computational expenses, ultimately reaching an equilibrium, as further demonstrated in the top subplot. Therefore, excessively high mining costs will dampen miners' incentives, which is detrimental to the security and stability of the blockchains.

\begin{figure}[t]
	\centering
	\includegraphics[width=7.8cm]{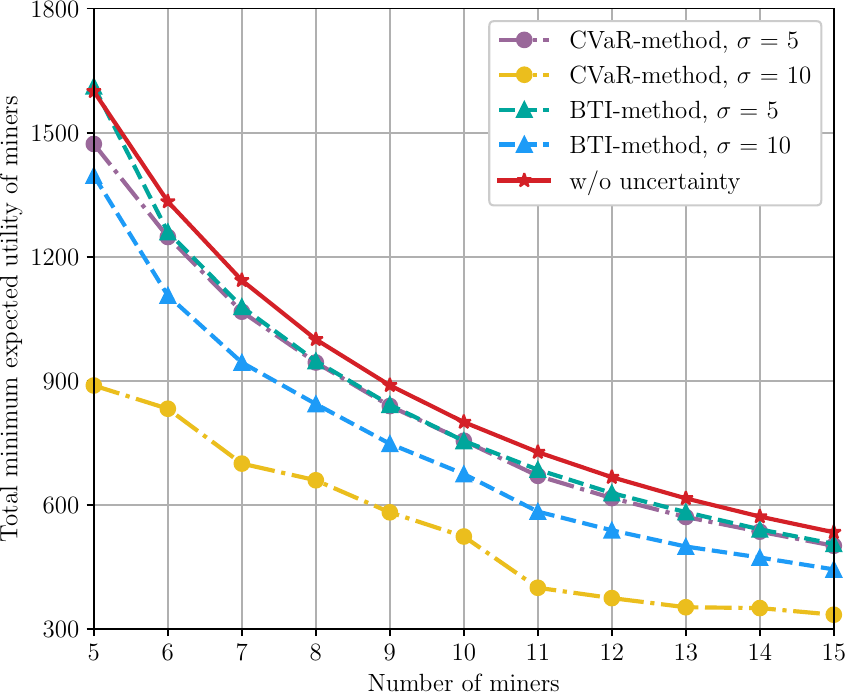}
	\caption{Performance comparison with different methods versus the number of miners under heterogeneous case.}
	\vspace{-4.5mm}
	\label{fig:num_he}
\end{figure}

In Figs.~\ref{fig:num_ho} and~\ref{fig:num_he}, we compare the optimization performance of different methods in blockchain networks with varying numbers of miners under homogeneous and heterogeneous conditions, respectively, to further demonstrate the scalability of the proposed approach. As observed, whether the scenario is homogeneous or heterogeneous, the total system utility gradually decreases as the number of miners in the network increases. First, we consider the homogeneous case: this occurs because, with a fixed mining reward, each miner's mining enthusiasm is the same. As the number of miners increases, the total computational power invested by the network rises, leading to higher computational costs, which in turn reduces system utility. Furthermore, as the number of miners continues to increase, the rate of decrease in system utility slows down. This is because, when the number of miners reaches a certain level, miners begin to reduce their computational contributions to mitigate the risks of the oversupply of miners relative to rewards scenario. Similarly, the same underlying reason applies in the heterogeneous case, but the variation in system utility is less smooth due to the differences in computing resources availability among miners.

\section{Conclusion} \label{sec:con}
In this paper, we have addressed the challenges arising from uncertain available computing resources in PoW blockchain mining. By formulating the problem as a distributionally robust optimization problem, we have provided a structured method to optimize mining strategies under resource uncertainty. Our framework, which employs the CVaR-based and BTI-based methods, effectively transforms the probabilistic constraints into tractable forms, enabling practical implementation in scenarios with unknown resource distributions. In environments with limited computing resources, miners often face substantial risks in terms of profitability. Our method provides a reliable mechanism to allocate resources while maintaining system stability. This contributes to the long-term sustainability and scalability of PoW blockchain systems, offering a promising tool for miners facing unpredictable computing environments.

\bibliographystyle{IEEEtran}
\bibliography{mainBib}

\end{document}